# Tunable band gaps and optical absorption properties of bent MoS$_2$ nanoribbons

Hong Tang[*], Bimal Neupane, Santosh Neupane, Shiqi Ruan, Niraj K. Nepal, and Adrienn Ruzsinszky

Department of Physics, Temple University, Philadelphia, PA 19122

**ABSTRACT**    The large tunability of band gaps and optical absorptions of armchair MoS$_2$ nanoribbons of different widths under bending is studied using density functional theory and many-body perturbation GW and Bethe-Salpeter equation approaches. We find that there are three critical bending curvatures, and the non-edge and edge band gaps generally show a non-monotonic trend with bending. The non-degenerate edge gap splits show an oscillating feature with ribbon width $n$, with a period $\Delta n = 3$, due to quantum confinement effects. The complex strain patterns on the bent nanoribbons control the varying features of band structures and band gaps that result in varying exciton formations and optical properties. The binding energy and the spin singlet-triplet split of the exciton forming the lowest absorption peak generally decrease with bending curvatures. The large tunability of optical properties of bent MoS$_2$ nanoribbons is promising and will find applications in tunable optoelectronic nanodevices.

Atomically thin two-dimensional (2D) layered materials, such as graphene and transition metal (di or mono) chalcogenides, are drawing a great attention in material science.[1-6] They are light weight and flexible, yet with a relatively high mechanical strength. They can be tailored into different shapes, intercalated by other atoms and molecules, strained in-plane, bent out-of-plane, rolled up into scrolls, wrinkled or folded in the 2D plane, and conformed onto a nanoscale-patterned substrate,[7] achieving varied, controllable properties. Additionally, they can be assembled, through interlayer van der Waals interactions,[8] into layer-on-layer stacked or twisted homo- or heterostructures, such as moire[9,10] patterned layered materials, leading to unprecedented, amazing properties. They are bestowed with a great promise in applications for next generation nanoelectronics[11] and optoelectronics.[4]

The reduced dimensionality in 2D materials usually results in reduced dielectric screening and enhanced electron-electron interactions,[12-15] and hence large exciton effects, which largely enhance the optical properties of 2D materials. Molybdenum disulfide (MoS$_2$) is a typical transition-metal dichalcogenide (TMD), featuring a high electron mobility comparable to graphene and a finite energy gap.[16] When decreasing from a bulk form down to a monolayer, MoS$_2$ crosses over from an indirect gap semiconductor to a direct one, as a result of inversion symmetry breaking in its honeycomb lattice structure.[2,12] Optical absorption and photoluminescence[12,17] determine the optical band gap of monolayer (1L) MoS$_2$ as 1.8~1.9 eV, while its fundamental band gap (or electronic band gap) is found to be ~2.5 eV by the delicate photocurrent[18] and scanning tunnelling spectroscopy[19] experiments, confirming the large binding energy (>570 meV) of the exciton in 1L MoS$_2$ systems. Qiu et al.[20] elaborately investigated the optical spectrum of 1L MoS$_2$ by using the GW+BSE[21] (Bethe-Salpeter equation) approach and revealed a large number and diverse character of bound excitons in it, suggesting its potential applications to electronics utilizing inter- and intraexcitonic processes.

---

[*] Corresponding author, email: hongtang@temple.edu

In terms of strain engineering,[22,23] it has been demonstrated that strain particularly plays an important role in manipulating the electronic and optical properties of graphene and TMDs. Strains change the relative positions of atoms, local potentials, and the orbital overlap between the metal and chalcogen atoms, resulting in significant alterations of electronic properties. At a homogeneous (or uniform) uniaxial strain ~2%, 1L MoS$_2$ changes from direct to indirect bandgap semiconductor, while with a biaxial tensile strain 10–15%, it undergoes a semiconductor-to-metal phase transition.[24-26] Since large homogeneous uniaxial or biaxial strains are relatively harder to realize in practical devices,[27] local nonuniform strains (LNS), by wrinkling,[23,28,29] indentation and interface conforming,[30,31] have been explored recently. Interestingly, LNS can generate novel effects, such as the exciton funnel,[23] in which before recombination excitons drift to lower bandgap regions caused by higher local strains, and spontaneous emission enhancement,[32] which is explored for ultracompact single-photon quantum light emitters.

Mechanical bending can provide effective LNS on nanoribbons, as shown by Yu et al.[33] and Nepal et al.[34] It was found that the bending induced shifting of edge bands and the charge localization of top valence bands can mitigate or remove the Fermi-level pinning and change the conductivity both along and perpendicular to the width direction of doped nanoribbons. Conduction along the width direction could be realized by attaching electrodes to the edges. Since edge band positions in band structures and edge band gaps are important for the optical absorption of nanoribbons, it is appealing to show how the edge bands will evolve with bending for varied widths of nanoribbons and how this will modify the optical properties. In this work, with density functional theory (DFT)[35-40] and many-body perturbation $G_0W_0$ computations,[41,42] we systematically investigate the band structures and band gaps of semiconducting armchair 1L MoS$_2$ nanoribbons with widths from 1.3 to 3.6 nm under different bending curvatures. It is found that the evolution of the edge gap has more features and shows a nonmonotonic trend with bending curvatures. The phenomenon is correlated with the complex strain patterns experienced in the bent nanoribbons. Furthermore, we use the GW+BSE approach to calculate the optical absorption spectra and reveal a large tunability of optical absorption by bending nanoribbons.

## Results and discussion

**Band gap tunability.** The nanoribbon (denoted as A$n$MoS$_2$) is formed by cutting from monolayer hexagonal MoS$_2$ and its two armchair edges are hydrogen passivated (Figures 1(a)-(c)). $n$ represents the number of MoS$_2$ units in one repeating unit along axis c. We study A$n$MoS$_2$ nanoribbons with $n$ from 9 to 24 and widths from 1.3 to 3.6 nm. The band structures of nanoribbons show a large tunability with bending. As shown for A13MoS$_2$ (Figure 1(d)), with an increase in bending curvature $\kappa$ ($\kappa = 1/R$, where R is an average curvature radius, Supporting Figure S1), the two nearly degenerate conduction bands C1 and C2 slowly approach to the Fermi energy, while the two valence bands V1 and V2 approach to the valence band continuum (VBC) and eventually merge into it at R = 10Å. C1, C2, V1 and V2 are mostly developed from the $d$ orbitals of the edge Mo atoms.[43] Both EG (edge band gap, see Figure 1(e) and caption) and NEG (non-edge band gap, Figure 1(f)) can change up to or over 50% with bending curvatures. There are three critical curvatures, namely $\kappa_0$, $\kappa_{c1}$ and $\kappa_{c2}$, dividing curvatures into four Regions I, II, III, and IV, shown in Figures 1(e) and (f). For A13MoS$_2$, $\kappa_0 = 0.04$/Å (R = 25Å), $\kappa_{c1} = 0.0625$/Å (R = 16Å) and $\kappa_{c2} = 0.100$/Å (R = 10 Å). With curvatures from zero to $\kappa_0$ (Region I), EG is almost unchanged, and NEG slightly increases. From $\kappa_0$ to $\kappa_{c1}$ (Region II), EG slightly decreases, while NEG increases further. From $\kappa_{c1}$ to $\kappa_{c2}$ (Region III), EG remains nearly constant. Region IV is for $\kappa > \kappa_{c2}$. Both in Regions III and IV, NEG decreases. At $\kappa_{c1}$, the border line between Regions II and III, NEG shows a hump (maximum). At $\kappa_{c2}$, V1 and V2 merge into VBC and EG turns to a quick decrease. All nanoribbons A$n$MoS$_2$ with $n$ from 9 to 23 show similar features (Supporting Figures S2-S3). However, with increasing width (larger $n$), all $\kappa_0$, $\kappa_{c1}$ and $\kappa_{c2}$ become smaller and Regions I-III gradually merge, and the NEG hump feature is reducing. Eventually, for A24MoS$_2$, EG shows nearly a constant behavior followed by a decrease with curvatures, while NEG shows a monotonic decrease. As shown in Figures 1(e) and (f), the trends of gap with curvatures from PBE,[35] SCAN,[36] TASK,[37] mTASK,[38] HSE06[39] and G$_0$W$_0$[41,42] are nearly the same, although the gaps themselves are not. PBE underestimates the band gaps, since it only has the ingredients of the local density and its gradient, without explicit inclusion of the nonlocal exchange effect, a factor important for an accurate description of band gaps. At meta-GGA level, SCAN can include some nonlocal exchange effects through the orbital dependent ingredient, and slightly improves the gaps. TASK with more nonlocality in the exchange part than SCAN, further improves the results, especially for EG. mTASK further increases the nonlocality in the exchange over TASK by lifting the tight upper bound for one- or two-electron systems and lowering the limit of the interpolation function $f_x(\alpha)$, resulting in a better description for low dimensional materials. For $d$-orbital MoS$_2$ nanoribbons, mTASK underestimates the gaps more than HSE06 does.

**Complex local strains in nanoribbons.** The strains in bending nanoribbons are highly non-uniform, complex, and closely related to the observed tunability of band gaps. For A13MoS$_2$, the strain in the xy-plane for the middle Mo atom layer (SXYM, Figure 2(a)) steadily decreases and gets more negative (compression) with R from ∞ to 16 Å, while the strain along the z direction for this middle Mo atom layer (SZM, Figure 2c) increases and gets more positive (tensile strains) with R from ∞ to 16 Å. After the critical $\kappa_{c1} = 0.0625/Å$ (R = 16Å), SXYM mostly increases (Figure 2b) and SZM mostly decreases (Figure 2d). The length of vector c of the supercell (LC) (Supporting Figure S4) of the nanoribbon also reaches the maximum at $\kappa_{c1}$, consistent with the turning of SZM at $\kappa_{c1}$. Since the relaxed structure of A13MoS$_2$ nanoribbon is nearly flat for R > 16 Å or $\kappa < \kappa_{c1} = 0.0625/Å$ (Supporting Figure S5), and SXYM is mostly negative for R > 16 Å, as in Figure 2(a), the nanoribbons experience a compression along the width direction, leading to a tensile expansion in the ribbon periodic direction, as expected for materials with a positive Poisson's ratio. So, LC increases up to $\kappa_{c1} = 0.0625/Å$ (R = 16Å). Furthermore, for R < 16 Å or $\kappa > \kappa_{c1}$, the nanoribbon begins to substantially bend outwards. The compression along the width starts to release and the nanoribbon is gradually getting tensile strains in the width direction, as in Figure 2(b) (the gradually positive strains of SXYM). At the same time, the expansion along the cell vector c starts to reduce and LC gradually becomes less and less.

The strain in the xy-plane for the outer S atom layer (SXYOS) and the strain along the z direction for the outer S atom layer (SZOS) (Supporting Figure S6) have the similar change trends with SXYM and SZM, respectively. The strain in the xy-plane for the inner S atom layer (SXYIS) is mostly compressive and shows mostly a decrease trend with $\kappa$, while the strain along the z direction for the inner S atom layer (SZIS) shares an approximately similar change trend with SZM, with more complex compressive and tensile patterns. For A$n$MoS$_2$ with even $n$, i.e., A12MoS$_2$ (Supporting Figure S7), the strains in the xy-plane (SXYM, SXYOS and SXYIS) have similar change trends to those of A13MoS$_2$, while the strains along the z direction (SZM, SZOS, and SZIS) have more complex patterns, due to the asymmetrical structures of the two edges. The narrow A11MoS$_2$ shows an additional feature. In Region III ($\kappa$ from 0.095 to 0.133/Å, Supporting Figures S3, S8), EG shows a sudden increase within the range of $\kappa$ from 0.1 to 0.125/Å. This relates to the special arrangement of edge atoms and strain patterns in the nanoribbon. For example, at $\kappa = 0.111/Å$ (R = 9 Å), the four Mo atoms on the two edges move along the z direction dramatically (Supporting Figure S9) and the strains associated with the four atoms are significantly large.

**Non-degenerate splits of edge bands.** The PBE value of NEG of flat nanoribbon shows a monotonic decrease with widths and almost reaches 1.7 eV (the PBE value of monolayer $MoS_2$) at $A24MoS_2$, as shown in Figure 3(a), while EG shows a vibrating feature with widths with a period $\Delta n = 3$ (Figure 3(b)). EG as a function of ribbon width under different curvatures is shown in Figures 3(c) and (d). As can be seen for both PBE and SCAN, under low curvatures ($\kappa \leq 0.04/\text{Å}$), EG still basically follows the same periodicity. However, for a large curvature, EG shows no obvious periodicity, although having a less regular oscillating feature. The bending space in nanoribbons will drastically alter the distribution and symmetry of wavefunctions, as well as the electronic band structures. For flat nanoribbons, the same period $\Delta n = 3$ is shown for the non-degenerate splits of edge bands (Figures 3(e) and (f)), namely, $\Delta E_C = E_{C2} - E_{C1}$, the energy difference between C2 and C1 at the $\Gamma$ point, and $\Delta E_V = E_{V1} - E_{V2}$, the energy difference between V1 and V2 at the $\Gamma$ point. For flat nanoribbons $AnMoS_2$ with $n = 9, 12, 15, 18, 21,$ and $24$ (i.e., $n = 3p$, where $p$ is an integer), both $\Delta E_C$ and $\Delta E_V$ are minimal, showing nearly degenerate edge bands. The flat nanoribbons with other $n$ values show non-degenerate edge bands around the $\Gamma$ point, with larger $\Delta E_C$ and $\Delta E_V$ for narrower nanoribbons (smaller $n$). For the same nanoribbon, $\Delta E_V$ is approximately three times $\Delta E_C$. Note that the flat nanoribbons with widths $n = 3p$ have larger EG values than those of neighboring $n$'s. This differs from the previous study[44] of $n = 3p - 1$ without hydrogen passivated nanoribbons. Hydrogen (H) passivation does not change the oscillating feature of EG with widths $n$, but shifts the maximum $n$'s by +1. As shown in Figure 3(g) for the flat $A13MoS_2$ nanoribbon, the distance u, which is the horizontal distance between the outmost Mo atom and the next outmost Mo atom, is approximately equal to the distance v, the horizontal distance between the outmost Mo and H atoms. For other flat $AnMoS_2$ nanoribbons, the situation is also similar. For non-passivated nanoribbons, the distance u is usually slightly less, since the outmost Mo atom will relax more towards the ribbon center. So, adding H atoms for a flat nanoribbon of width $n$ will make its effective width increased (Supporting Figure S10) and approximately equal to the width $m$ of the nanoribbon without H passivation, where $m = n + 2$. The H passivated nanoribbon of width $n = 3p$ has a high EG value, so does the unpassivated nanoribbon of width $m$. Since $m = n + 2 = 3p + 2 = 3(p + 1) - 1 = 3p' - 1$, where $p'$ is also an integer. This basically explains the "-1" difference in the maximum $n$'s for nanoribbons with and without H passivation.

Figure 4 shows the nondegenerate splitting of the lower ($\Delta E_V$) and the upper ($\Delta E_C$) edge band gap as a function of bending curvature for different nanoribbon width from SCAN. As can be seen, the splits $\Delta E_C$ and $\Delta E_V$ are robustly kept when bending is applied. For example, for $A13MoS_2$, the split $\Delta E_V$ is kept from $\kappa = 0$ to $0.091/\text{Å}$, until V1 and V2 merge into VBC at $\kappa = 0.10/\text{Å}$. Bending can even enhance the split $\Delta E_V$ ($\Delta E_C$), as can be seen for $A11MoS_2$ ($A9MoS_2$) from $\kappa = 0$ to $0.091/\text{Å}$ (0 to $0.10/\text{Å}$), and induce the split $\Delta E_V$, as can be seen for $A12MoS_2$ from $\kappa = 0$ to $0.10/\text{Å}$. The splits $\Delta E_C$ and $\Delta E_V$ on flat nanoribbons may be due to the quantum confinement effect along the width direction. Bending can couple with the quantum confinement effect and enhances the splits, especially for narrower nanoribbons.

In Figure 4, for about $\kappa \leq 0.02/\text{Å}$, the ordering of the curves for different width $n$ is basically the same as that of the flat case, implying that $\Delta E_V$ and $\Delta E_C$ still have the same periodicity and oscillating behavior as that of the flat nanoribbons. When increasing curvatures, the ordering of the $\Delta E_V$ curves for wider nanoribbons ($n$ from 18 to 24) is also approximately unchanged, before the lower edge bands merge into the valence continuum at about $\kappa = 0.07/\text{Å}$. However, the orderings of the $\Delta E_V$ and $\Delta E_C$ curves for narrower nanoribbons ($n$ from 9 to 17) all show a more complex feature with larger curvatures. This is echoed with the complex changing feature of EG vs. large curvatures (supporting Figure S2). In those large curvature regions, $\Delta E_V$ and $\Delta E_C$, especially for narrow nanoribbons, will not follow the same periodicity and oscillating behavior as that of the flat ones.

**Tunable optical properties.** The complex strain patterns in the bent nanoribbons control the varying features of band gaps. Those features will result in varying exciton formations and optical properties. To elucidate the tunable optical absorptions of bent nanoribbons, we calculate the absorption spectra with GW+BSE for nanoribbons with bending curvatures in the different curvature regions. The calculated optical absorption spectra of A13MoS$_2$ nanoribbon are shown in Figure 5, showing a large tunability of absorption with bending curvatures. The absorption peaks are generally shifted to lower energies and more absorption peaks occur. For example, there is almost no absorption at photon energies 0.35 and 1.25 eV for a flat ribbon, while the bent nanoribbon with R = 6 Å produces a weak absorption peak at 0.35eV and a strong peak at 1.25 eV. With increasing curvatures, the quasiparticle gap of the nanoribbon decreases (Table 1), more and more bright exciton states appear within the energy range below the fundamental gap, and the bending nanoribbon shows a broad absorption within this energy range.

At R = ∞ (flat), peak A′ consists of two degenerate (energy difference < 3meV) exciton states at energy 0.57 eV, and each of them is due to a combination of four transitions, namely, V1 to C1, V1 to C2, V2 to C1, and V2 to C2, mainly around the Γ point. The wavefunction of the exciton is shown in Figure 6. The binding energy of these excitons is 1.41 eV, which is bigger than that (0.96 eV [20]) of the lowest exciton in 1L MoS$_2$, due to the further reduced screening and enhanced electron-electron interaction in the one-dimensional nanoribbon. Peak B′ mainly consists of two degenerate exciton states at 0.94 eV. One is mainly due to the transition around Γ from V3 to C1, and the other from V3 to C2. There are two degenerate and relatively weak exciton states at 1.01 eV, merged in the right bottom of peak B′. They are the excited states of the excitons forming peak A′. Peak C′ consists of two sets of exciton sates. The first set, with an energy at 1.11 eV, is two degenerate exciton states mainly corresponding to the transition V4 to C1 (or V4 to C2) around Γ, while the second set, with a slightly higher energy at 1.17 eV, is two degenerate exciton states mainly due to V3 to C1 (or V3 to C2) around k points slightly away from, but near Γ. The wavefunction of this second set exciton has a nodal feature, indicating that it is the excited state of the exciton forming peak B′ at 0.94 eV. Peaks D′ and E′ are relatively weak and mainly due to transitions from lower valence bands (V5 or V6) to upper edge bands (C1 or C2), and their wavefunctions in k space have more complex nodal features. Peak F′ is mainly due to transition V1→C3 around Γ. This exciton has an energy very close to the fundamental gap and hence a small binding energy of 0.09 eV.

At R = 13 Å, the excitonic composition of peak A″ is similar to that of peak A′. However, due to bending, the two originally degenerate exciton states red shift to 0.5 and 0.55 eV, respectively. This makes the overall peak A″ broadened and located at a lower energy than peak A′. Peak B″ becomes complex. Overall, it is still similar to peak B′, and consists of exciton states corresponding to transitions around Γ for V3→C1 (or V3→C2), and the excited states of excitons forming peak A″. The overall peak position of B″ red shifts, compared to peak B′. Similarly, peaks C″ to G″ are complex and involve mixtures of transitions involving lower valence bands (V3 to V6) and upper edge bands (C1 and C2).

At R = 9 Å, peaks A‴ and B‴ are, respectively, due to two excitons, each of which is mainly from the mixed transitions involving V2→C1, V3→C1, V2→C2, and V3→C2 around Γ. The exciton corresponding to peak A‴ is at 0.5 eV, while the one for B‴ is at 0.57 eV, and the energy difference may be mainly due to bending effects. Peaks A‴ and B‴ are related to, but different from peaks A′ and A″. At R = 9 Å, the lower edge bands, named as V1 and V2 before merging into VBC and named as V2 and V3 after the merging, are already merged into VBC. As can be seen (Figure 5(k)), V1, V2 and V3 are very close to each other around Γ, and the transition V1→C1/C2 has very small contributions (<4%) to peaks A‴ and B‴. In fact, the exciton states corresponding to transitions V1→C1/C2 are moved to 0.65 and 0.7 eV, and are relatively weak and embedded in the valley between B′′′ and C′′′. Yu et al. [33] showed that the charge density of the continuum valence band maximum (V3 at Γ before the merging) changes from a nearly uniform and symmetric distribution along the nanoribbon width for flat case to a nonuniform distribution concentrating in the middle region of the nanoribbon for large bending curvatures, resulting in an asymmetric wavefunction over the inner and outer sulfur layers. This may also change the parity of the wavefunction. On the other hand, both the upper and lower edge bands are located mainly near the two edges, thus less affected by the bending. The symmetry and parity of wavefunctions involved is important to the formation of optically active bright excitons.[45] So, the band inversion of V1/V2 with V3 around Γ at R = 9 Å may change the symmetry and parity of wavefunction of the continuum valence band maximum, in a manner less favoring a large oscillator strength for the optical transition, leading to a weak absorption contribution from the transition V1→C1/C2. Peaks C′′′, D′′′ and E′′′ form a broad composite peak and consist of many exciton states involving mixed transitions from V1, V2 and V3 to C1 and C2. Some of these excitons bear features of the excited states of the excitons forming peaks A′′′ and B′′′. Peaks F′′′ and G′′′ are weak and involve transitions from lower valence bands (V4, V5, and V6) to C1 and C2. Peak H′′′, similar to peak F′, is mainly due to exciton states involving transitions from V1, V2, V3 to C3, since V1, V2, and V3 are very close around Γ. The binding energy of the exciton states mainly contributing to peak H′′′ is 0.24 eV, larger than that for peak F′.

At R = 6 Å, the main contribution to peak A′′′′ is two degenerate excitons at 0.33 eV and they are mainly due to transitions V1/V2→C1/C2 around Γ. This is mainly due to the closeness of V1 and V2 around Γ, and V3 is lower than V1 and V2 there (Figure 5(l)). Note that the height of peak A′′′′ is apparently much lower than other main peaks. Peak B′′′′ is mainly due to excitons of transitions V3→C1/C2 and V4→C1/C2. Peaks from C′′′′ to I′′′′ become more complex. The bending lowers EG, and the conduction band continuum (CBC) also shifts downwards. This makes the upper edge bands C1 and C2 closer to CBC, and more transitions relating lower valence bands (V3, V4, V5, etc.) and higher conduction bands (C3, C4, etc.) occur. For example, peak G′′′′ has contributions of excitons involving mixed transitions from V3-V5 to C1-C4.

Figures 5(e)-(h) show the exciton energy spectra under different curvatures, with no feature of the 2D hydrogenic model,[46] due to the varying dielectric screening effect in the confined and layered nanoribbon structures. The A12MoS$_2$ nanoribbon also has similar tunable properties (Supporting Figure S21). With increasing curvature, the main absorption peaks shift to lower frequencies and become broader. Bending activates more exciton states and many of them contribute to the optical absorptions. The lowest energy exciton's binding energy generally decreases with bending (Table 1), basically consistently with an increase in the static dielectric constant (Supporting Figures S22 and S23) with bending, while the wavefunction distortions induced by the bending surface (or space) in bent nanoribbons may also influence the binding. The spin singlet-triplet splitting of the lowest energy exciton is about 0.22 eV for flat ribbons and shows a tunability (decrease) with bending (Table 1). The bending space reduces the overlap between electron and hole wave functions and thus results in a decrease in e–h exchange interactions. We think that the edge states in the nanoribbons have similar properties to those of the defect states[47] in monolayer TMDs and the relatively large and tunable singlet-triplet splitting can make the nanoribbon system suitable for quantum information applications.[48] Figure 7 shows the energy levels of several low energy singlet and triplet excitons for the A13MoS$_2$ nanoribbon at three bending curvatures. Since the typical lifetime of an exciton in monolayer MoS$_2$ can be about 10 nanoseconds,[49] it is feasible to optically excite the system from one triplet state to another in the bent nanoribbon. Also, the intersystem crossing in MoS$_2$ system is realizable, evidenced by the phosphorescence application of MoS$_2$ quantum dots,[50] and it becomes more realizable in bent nanoribbons since bending usually increases the spin-orbit coupling, which assists in changing the spin

during the intersystem crossing process. Besides, by appropriate doping or introducing defects, one may realize combination or hybridization between localized states and edge states in the nanoribbons, creating possible novel quantum states.

The band structures of the A12MoS$_2$ and A13MoS$_2$ nanoribbons with SOC are also calculated (Supporting Figure set SA). It shows that there is a negligible difference in the band structures between those with SOC and without SOC, before the lower edge bands merge into the valence continuum. Otherwise, there are small SOC induced splitting (~50 meV) in the top valence bands at k points away from Γ. Since the SOC induced splitting is much smaller than the quasiparticle fundamental gaps (~ > 1.5 eV) and not located at Γ, this will have very limited influence on the calculated optical absorption. However, even the small SOC effects may have an important consequence to the dynamic process involving excitons, such as the intersystem crossing.

The large tunability of electron energy loss spectrum (EELS) and absorption coefficient $\alpha$ with bending is shown in Figure 8. As can be seen, both EELS and $\alpha$ are very low at energy around 1.5 eV for low bending curvatures, while they are large at large curvatures. Since the electron energy loss spectrum and the optical absorption coefficient are the quantities conveniently accessible by experiments, this can be utilized as a means to detect or control the bending for nanoribbon devices. Also, the data presented in this work can serve as a guide for the future experiments. As shown here, the mechanical bending is an effective means to control and fine tune the optical properties of nanoribbons.

## Conclusion

In conclusions, from first-principles calculations DFTs and GW+BSE, we assessed the large tunability of band gaps and optical absorptions of armchair MoS$_2$ nanoribbons of different widths under different bending curvatures. We find that there are three critical bending curvatures $\kappa_0$, $\kappa_{c1}$ and $\kappa_{c2}$ with $\kappa_0 < \kappa_{c1} < \kappa_{c2}$. Below $\kappa_0$, the edge gap is almost unchanged, while from $\kappa_0$ to $\kappa_{c1}$, it slightly decreases. The non-edge gap slowly increases from zero curvature to $\kappa_{c1}$. From $\kappa_{c1}$ to $\kappa_{c2}$, the edge gap nearly keeps constant, and beyond $\kappa_{c2}$, it decreases. From $\kappa_{c2}$ and on, the lower edge bands merge into the valence band continuum. The non-edge gap decreases from $\kappa_{c1}$ and on, and has a maximum around $\kappa_{c1}$, consistent with the maximum strain along the ribbon length direction around $\kappa_{c1}$. For wider nanoribbons (width = 3.6 nm), Regions I, II and III merge, and the non-edge gap shows a monotonic decrease with bending, while the edge gap shows a constant behavior followed by a decrease under increasing curvatures. The edge gaps and the non-degenerate edge gap splits show an oscillating feature with ribbon width $n$, with a period $\Delta n = 3$, due to quantum confinement effects. The non-degenerate edge gap splits generally persist with bending. Bending generally induces more exciton states and they contribute to controllable optical absorptions. The induced excitons are related to the bands near the gap and the subtle changes of these bands with bending. The binding energy and the spin singlet-triplet split of the exciton forming the lowest absorption peak generally decreases with bending curvatures. This latter phenomenon opens opportunities for bent nanoribbons to utilize excitons in quantum information science. Since MoS$_2$ nanotubes have already been synthesized,[51,52] it may be feasible to realize the bent nanoribbons by embedding MoS$_2$ nanotubes into nano-troughs and etching or eroding out some parts of the nanotubes. The large tunability of optical properties of bending MoS$_2$ nanoribbons is appealing and will find applications in tunable optoelectronic nanodevices.

## Methods

**Computational details for band gaps and optical absorption.** Density functional theory (DFT) calculations were conducted in the Vienna Ab initio Software Package (VASP)[40] with projector augmented-wave pseudopotentials.[53,54] PBE,[35] SCAN,[36] TASK,[37] mTASK,[38] HSE06[39] approximations were used to calculate the band structures of nanoribbons. The vacuum layer of more than 12 Å is added along the direction of nanoribbon width and inserted along the direction perpendicular to the 2D surface of the nanoribbon, to avoid the interactions between the nanoribbon and its periodic images. The energy cutoff is 500 eV. The k-point mesh of $1 \times 1 \times 8$ was used for all nanoribbons. All nanoribbons were fully structurally relaxed with PBE with all forces less than 0.008 eV/Å. During the relaxation, the x and y coordinates of the two outer most metal atoms on the two edge sides were fixed, while their coordinates along the ribbon axis direction, which is the z direction, and all the coordinates of other atoms were allowed to relax. The $G_0W_0$[41,42] and $G_0W_0$+BSE[21] calculations were conducted in BerkeleyGW[41] by pairing with Quantum ESPRESSO[55]. The wavefunction energy cutoff is 65 Ry (~880 eV). The energy cutoff for the epsilon matrix is 20 Ry (~270 eV). The k-point mesh of $1 \times 1 \times 32$ and both valence and conduction bands of 6 was set for optical absorption calculations. The band number for summation is 1000. The correction of the exact static remainder and the wire Coulomb truncation for 1D systems were also used.

**Calculation methods for strains.** As shown in Figure 2(e), Position 1 in "Position along the x direction" represents the atom pair of Mo1 and Mo2. Position 2 represents the pair of Mo2 and Mo3, and so on for other positions. For Position 1, the strain in the xy-plane for the middle Mo atom layer (SXYM) is calculated as $100 \times (l_{xy} - l_{xy}^0)/l_{xy}^0$, where $l_{xy} = \sqrt{(x_{Mo1} - x_{Mo2})^2 + (y_{Mo1} - y_{Mo2})^2}$, and the strain along the z direction for this middle Mo atom layer (SZM) is calculated as $100 \times (l_z - l_z^0)/l_z^0$, where $l_z = \sqrt{(z_{Mo1} - z_{Mo2})^2}$. Both $l_{xy}^0$ and $l_z^0$ are the corresponding values taken from a reference monolayer MoS$_2$, which is relaxed with PBE with the relaxed lattice constants $a_1 = 3.182$Å and $a_3 = 3.127$Å, where $a_1$ is the distance between two adjacent Mo atoms and $a_3$ is the distance between the two nearest S atoms from the upper and lower S layers. The strains at other positions, as well as for S atom pairs in the out or inner S atom layer, are calculated in the similar way.

**Supporting Information**

Evolutions of NEG and EG for A$n$MoS$_2$ nanoribbons with *n* from 9 to 24 with bending curvatures; the evolution of strains with curvature radii for the A11MoS$_2$, A12MoS$_2$, and A13MoS$_2$ nanoribbons; the length of vector c of the supercell as a function of curvatures for nanoribbons A$n$MoS$_2$ with *n* form 9 to 24; plots of exciton peaks for A13MoS$_2$ nanoribbon; optical absorption spectra of A12MoS$_2$ nanoribbon under different curvature radii; the modulus square of the wavefunction of the C1 and V1 bands of A13MoS$_2$ nanoribbon; real part of dielectric functions, EELS and absorption coefficients of A12MoS$_2$, A13MoS$_2$ nanoribbons under different curvature radii; Band structures of A12MoS$_2$ and A13MoS$_2$ nanoribbons with PBE and PBE+SOC.

**Acknowledgement**

This material is based upon work supported by the U.S. Department of Energy, Office of Science, Office of Basic Energy Sciences, under Award Number DE-SC0021263. This research used resources of the National Energy Research Scientific Computing Center, a DOE Office of Science User Facility supported by the Office of Science of the U.S. Department of Energy under Contract No. DE-AC02-05CH11231.

## Author contributions

H.T. and A.R. designed the research. H.T. B.N. and S.N. performed the computations. H.T., B.N., N.K.N. and A.R. contributed to analyses of results. S.N. and S.R. contributed to discuss the results. H.T wrote the manuscript with contributions from N.K.N. and A.R. All authors reviewed the manuscript.

## Competing interests

The authors declare no competing interests.

## Data availability

The data that support the plots within this paper are available from the corresponding authors upon request.

## Code availability

The code for the calculations of bent coordinates and strains of nanoribbons described in this paper are available from the corresponding authors.

Table 1. The quasiparticle gap $E_g$, the energies of exciton forming the lowest energy peak in the absorption spectra $E_A$ (for spin singlet), $E_A^{triplet}$ (for spin triplet), the binding energy of this exciton $E_b$ (for spin singlet), and the spin singlet-triplet split $\Delta^{S-T}$ for A12MoS$_2$ and A13MoS$_2$ nanoribbons under different bending curvature radii. $E_g$ is from GW calculations. $E_b = E_g - E_A$ and $\Delta^{S-T} = E_A - E_A^{triplet}$. Energy unit in eV.

| | A13MoS$_2$ | | | | |
|---|---|---|---|---|---|
| | $E_g$ | $E_A$ | $E_b$ | $E_A^{triplet}$ | $\Delta^{S-T}$ |
| R = ∞ (flat) | 1.98 | 0.57 | 1.41 | 0.35 | 0.22 |
| R = 13 Å | 1.91 | 0.50 | 1.41 | 0.30 | 0.20 |
| R = 9 Å | 1.88 | 0.50 | 1.38 | 0.30 | 0.20 |
| R = 6 Å | 1.47 | 0.33 | 1.14 | 0.28 | 0.05 |

| | A12MoS$_2$ | | | | |
|---|---|---|---|---|---|
| | $E_g$ | $E_A$ | $E_b$ | $E_A^{triplet}$ | $\Delta^{S-T}$ |
| R = ∞ (flat) | 2.06 | 0.59 | 1.47 | 0.37 | 0.22 |
| R = 14 Å | 1.89 | 0.49 | 1.40 | 0.30 | 0.19 |
| R = 10 Å | 1.85 | 0.45 | 1.40 | 0.26 | 0.19 |
| R = 7 Å | 1.66 | 0.41 | 1.25 | 0.24 | 0.17 |

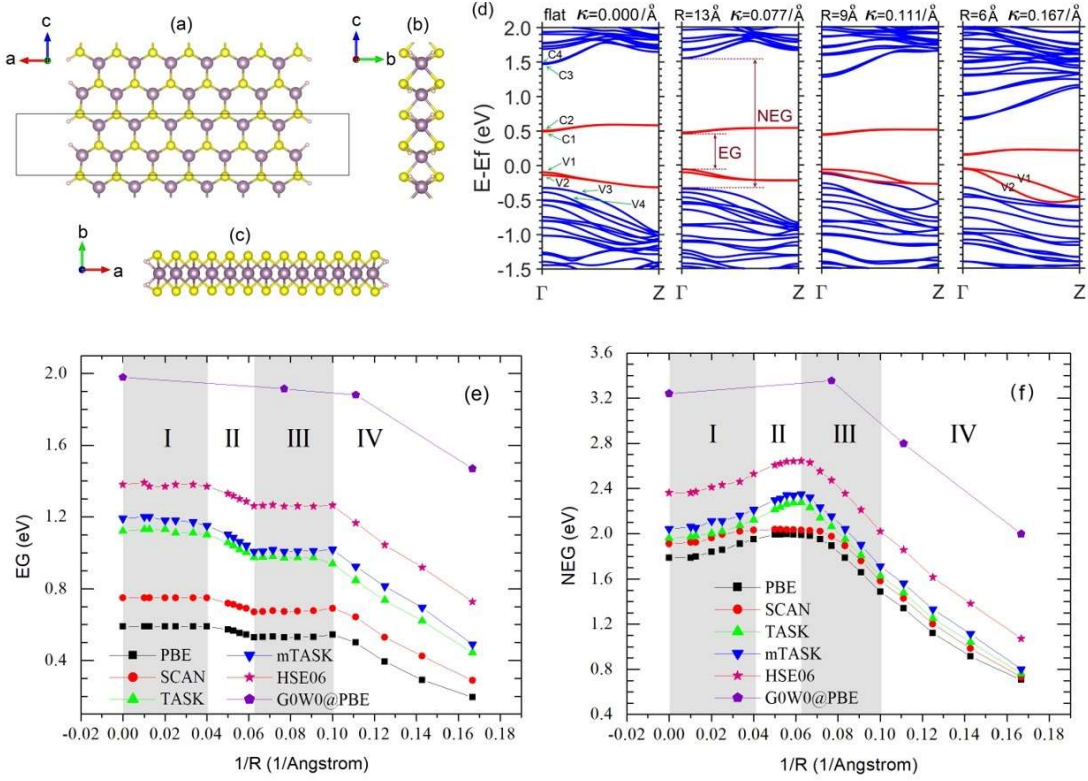

Figure 1. The structure of hexagonal armchair monolayer MoS$_2$ nanoribbon A$n$MoS$_2$ with $n$=13 and its band structure and gap evolutions with the bending curvature radius. (a) structure view along axis y. The supercell vectors a, b and c are aligned with axes x, y and z, respectively. The box outlines the periodical unit of the nanoribbon along axis z. The two hydrogen-passivated armchair edges are on the left and right sides. The view along axis x is in (b) and along axis z is in (c). The blue balls represent Mo atoms, the yellow ones for S atoms, and small white ones for H atoms. The PBE band structure evolution with the bending curvature radius is in (d). The conduction bands C1, C2, C3… are numbered upwards, while the valence bands V1, V2, V3… are numbered downwards. The four bands near the Fermi level are plot in red, while others are in blue. The edge band gaps (EG) in (e) and non-edge band gaps (NEG) in (f) as a function of bending curvature $\kappa$ for the A13MoS$_2$ nanoribbon are shown. The shaded areas highlight the different curvature regions. The results from PBE, SCAN, TASK, mTASK, HSE06 and G$_0$W$_0$ are shown, and have nearly the same trend. EG is defined as the energy difference between C1 and V1 at the Γ point, and NEG as that between C3 and V3. When V1 and V2 are merged into the valence band continuum, NEG is the difference between C3 and V1.

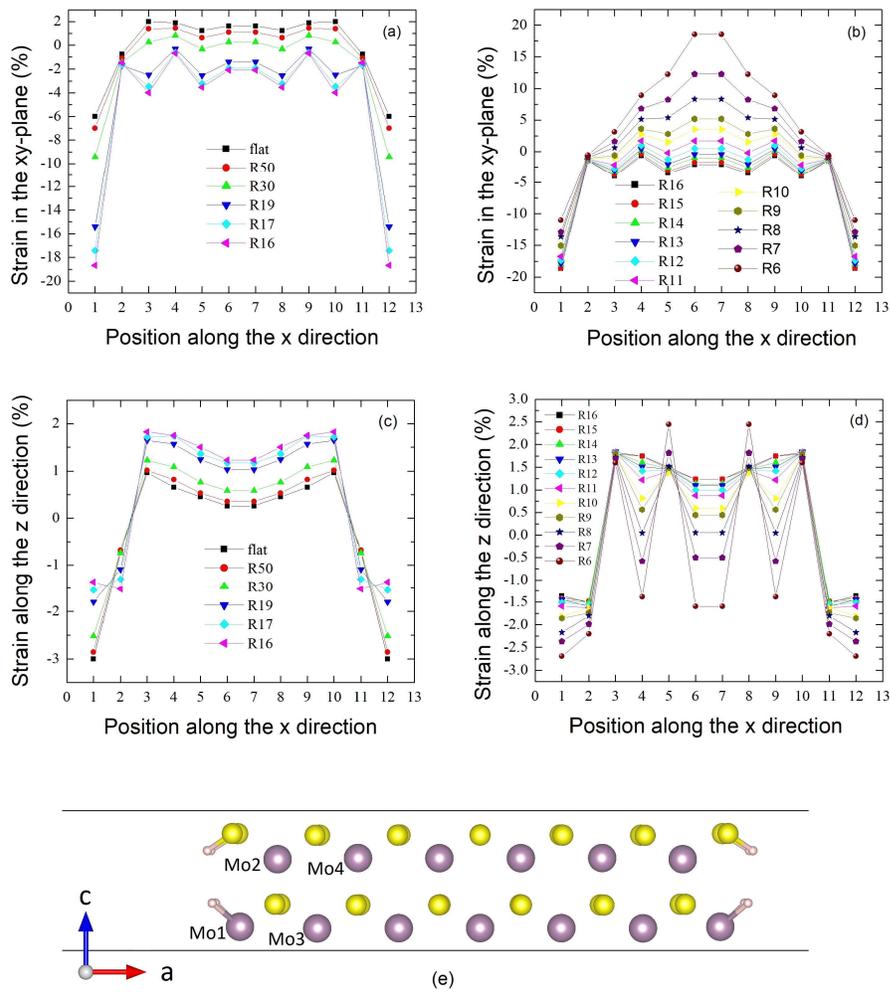

Figure 2. The evolution of strains with curvature radii for the A13MoS$_2$ nanoribbon and the schematic showing the calculation of the strains. The strain in the xy-plane for the middle Mo atom layer (SXYM) with R from ∞ (flat) to 16 Å in (a), and with R = 16Å to 6 Å in (b). In (a) SXYM steadily decreases and gets more negative with R from ∞ to 16 Å. In (b) SXYM mostly increases beyond the critical $\kappa_{c1} = 0.0625/Å$ (R = 16Å). R50 represents R = 50Å and so on. The strain along the z direction for this middle Mo atom layer (SZM) with R from ∞ (flat) to 16 Å in (c), and with R = 16Å to 6 Å in (d). The schematic graph showing the atomic positions in a nanoribbon is in (e). The details for the positions along the x direction and the calculation of strains are described in the Methods section.

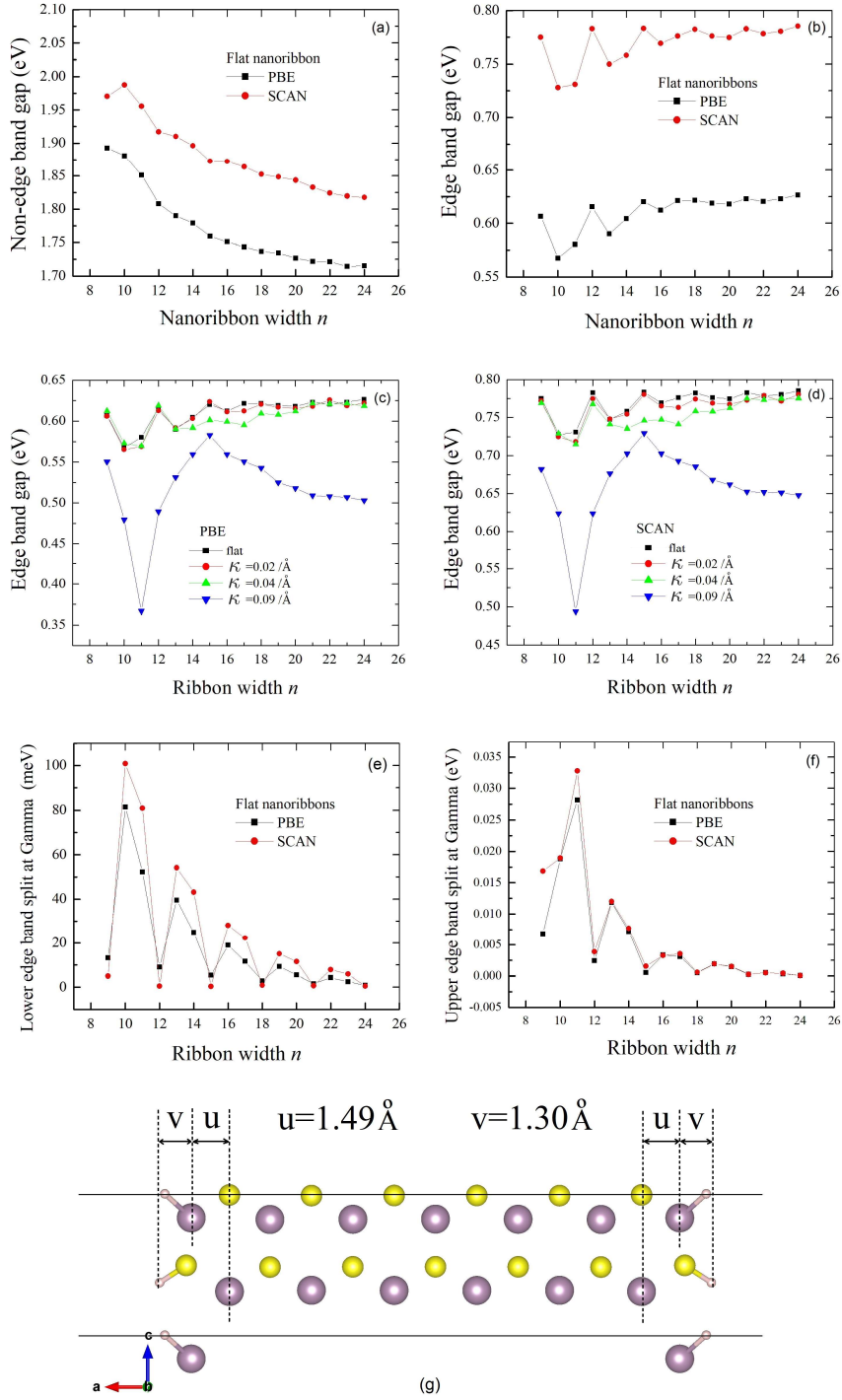

Figure 3. The gaps of nanoribbons A$n$MoS$_2$ with $n$ from 9 to 24, obtained with the PBE and SCAN functionals, as the function of ribbon width and the relaxed atomic structure of the flat A13MoS$_2$ nanoribbon. The non-edge band gap (NEG) is in (a), the edge band gaps (EG) in (b), (c) and (d), the non-degenerate splits of edge bands $\Delta E_V$ (lower edge band split) in (e) and $\Delta E_C$ (upper edge band split) in (f). In (c) and (d), the curves under different bending curvatures are shown. NEG approximately shows a monotonic decrease trend with $n$, while EG shows an oscillating feature with a period $\Delta n = 3$. $\Delta E_C =$

$E_{C2} - E_{C1}$, defined as the energy difference between C2 and C1 at the Γ point (see Figure 1d), and $\Delta E_V = E_{V1} - E_{V2}$, the energy difference between V1 and V2 at the Γ point. In (g), the distance u, which is the horizontal (ribbon width direction, or cell-vector a direction) distance between the outmost Mo atom and the next outmost Mo atom, is approximately equal to the distance v, the horizontal distance between the outmost Mo and H atoms.

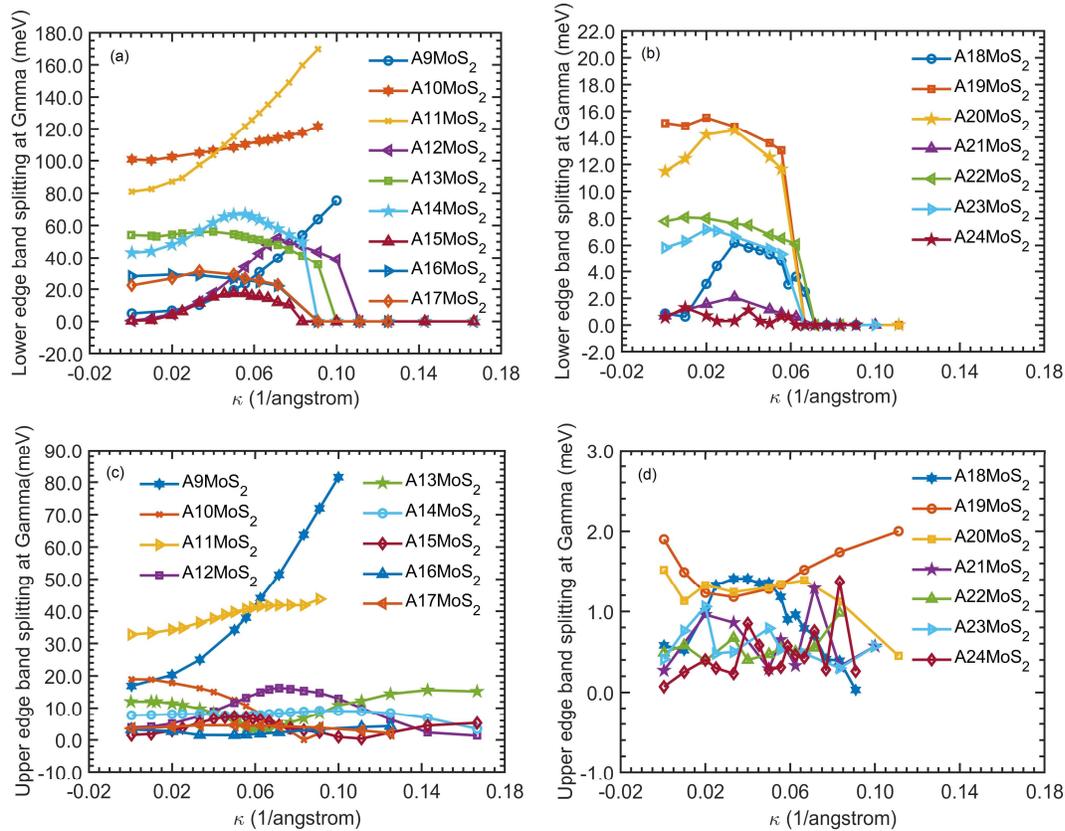

Figure 4. The nondegenerate splitting of the lower edge band gap ($\Delta E_V$) and the upper edge band gap ($\Delta E_C$) of A$n$MoS$_2$ nanoribbons as a function of bending curvature $\kappa$ calculated from SCAN for different nanoribbon width $n$. The lower edge band gap splitting is in (a) $n$ from 9 to 17 and (b) $n$ from 18 to 24. The upper edge gap splitting is in (c) $n$ from 9 to 17 and (d) $n$ from 18 to 24.

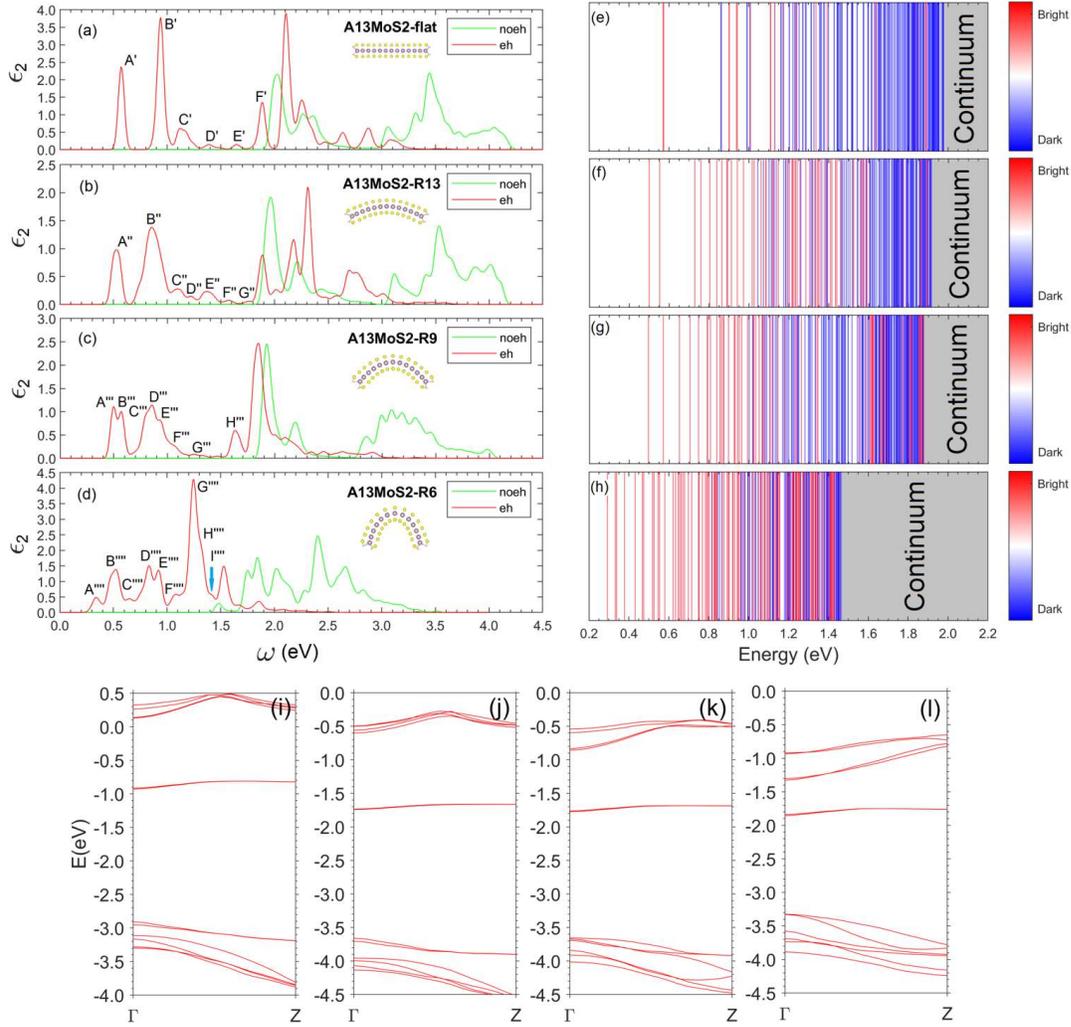

Figure 5. The optical absorption spectra, the corresponding exciton spectra, and the GW band structures of A13MoS$_2$ nanoribbon under different bending curvatures. The optical absorption spectra are plotted as the imaginary part of the dielectric function as a function of photon energy for curvature radii (a) R = ∞ (flat), (b) R = 13Å, (c) R = 9Å, and (d) R = 6Å. Red curves represent the GW+BSE results with electron-hole (eh) interactions and the green ones are for the results without eh (noeh) interactions, both with constant broadening of 26meV. Labels in (a)-(d), i.e., A′, A″, A‴, A⁗, B′, etc. represent different peaks. Inset graphs in (a)-(d) show the structures of bent nanoribbons. The exciton spectra show the energy positions of exciton states for curvature radii (e) R = ∞ (flat), (f) R = 13Å, (g) R = 9Å, and (h) R = 6Å. Bright (dark) exciton states are represented by red (blue) lines. The GW band structures are in (i) R = ∞ (flat), (j) R = 13Å, (k) R = 9Å, and (l) R = 6Å.

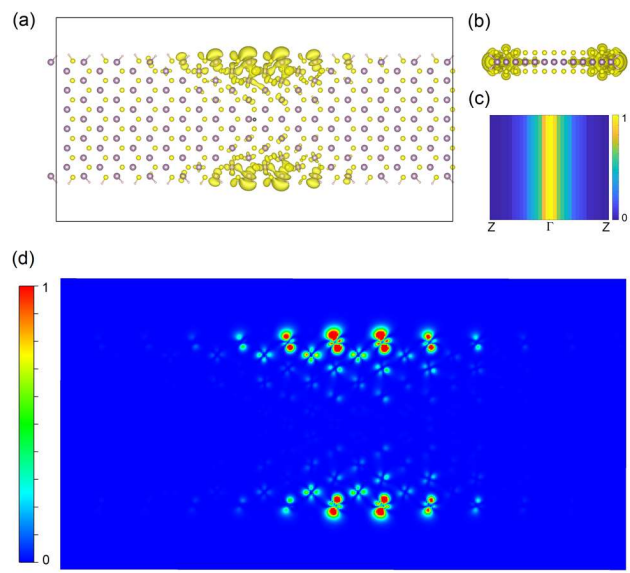

Figure 6. The exciton state forming peak A′ (Figure 5a) in real space (a) top view, (b) side view, and in k space (c). The hole (black spot in (a)) is located at the center of the ribbon and near a Mo atom. The isosurface contour of the modulus squared exciton wavefunction is shown in (a) and (b). The profile (arbitrary unit) of the modulus squared exciton wavefunction in k space is shown in (c), showing it is around the Γ point, and in real space is shown in (d), showing the electron in the exciton is mainly located on the edge Mo atoms. The plots for other exciton states are in Figures S11-S20 (Supporting Information).

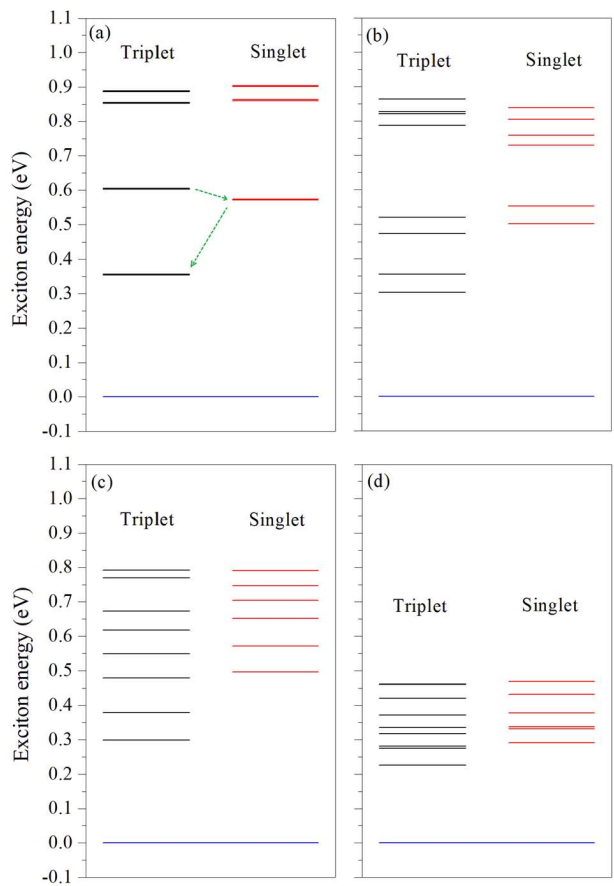

Figure 7. Energy levels of several low energy singlet and triplet excitons for A13MoS$_2$ nanoribbon at different bending curvature radii R: (a) R = ∞, (b) R = 13Å, (c) R = 9Å and (d) R = 6Å. The dashed green arrows in (a) indicate the possible intersystem crossing between singlet and triplet states.

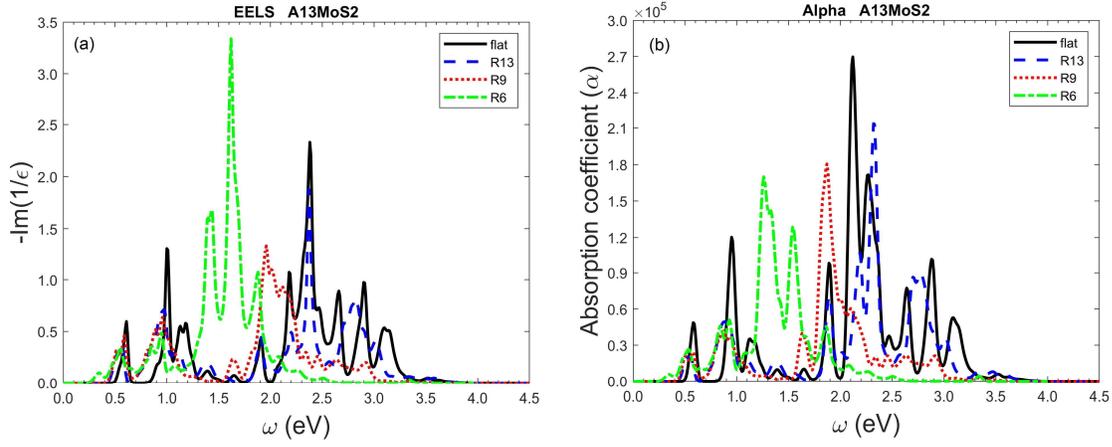

Figure 8. The electron energy loss spectrum (EELS) in (a) and absorption coefficient in (b) of A13MoS$_2$ nanoribbon under different bending curvatures. EELS is calculated as the imaginary part of $-1/(\varepsilon_1 + i\varepsilon_2)$. The absorption coefficient $\alpha$ is calculated as $\omega\varepsilon_2/(nc)$ (cm$^{-1}$, Gaussian unit), where $n = \sqrt{(\sqrt{\varepsilon_1^2 + \varepsilon_2^2} + \varepsilon_1)/2}$ is the refractive index and $c$ is the speed of light in vacuum. R13 represents the bending curvature radius R $= 13$Å and so on. The plots for the A12Mos$_2$ nanoribbon are in Figure S24 (Supporting Information).